# Fermi surface of yttrium

G. Kontrym-Sznajd[1], H. Sormann[2] and R. N. West[3]

[1] W.Trzebiatowski Institute of Low Temperature and Structure
Research, Polish Academy of Sciences, P.O.Box 1410, 50-950 Wrocław 2, Poland

[2] Institut für Theoretische Physik, Technische Universität Graz, Petersgasse 16,
A-8010 Graz, Austria

[3] Department of Physics, University of Texas at Arlington, P. O. Box 19059,
Arlington, Texas 76019, USA

**Abstract.** Electron-positron momentum densities in Y, reconstructed from two-dimensional angular correlation of annihilation radiation spectra, are compared with the theoretical predictions of fully-relativistic augmented plane-wave calculations. Knowledge of the theoretical densities and of the effects on them of certain symmetry selection rules has allowed us to separate two hole Fermi surfaces in the third and fourth bands and to establish some Fermi momenta for each of them.

Almost all theoretical calculations of electronic structure of trivalent yttrium [1] deliver the same Fermi surface (FS) topology: The first two bands are fully occupied and the FS exists in the 3$^{rd}$ and 4$^{th}$ bands. Experimental data was initially scarce. An early, one-dimensional angular correlation of annihilation radiation (1D ACAR) study [2], disclosed little about the FS but, some ten years later, a dHvA experiment [3] provided a few parameters in qualitative agreement with band structure calculations by Loucks [1a]. A recent [4] k-space analysis [5] of 2D ACAR data exploited the near coincidence of the 3$^{rd}$ and 4$^{th}$ band surfaces on the KMLH Brillouin zone face. By that means the authors obtained the shape and size of the so-called *webbing* feature at that face, but were, nevertheless, unable to provide any information about the individual surfaces there or elsewhere. Here we remedy that deficiency in a further interpretation of the same 2D ACAR data but, this time, in terms of the full 3D electron-positron momentum density, $\rho(p)$, and in a comparison with theoretical, fully relativistic, APW results.

As in the earlier work [4] we have reconstructed the experimental momentum density $\rho(p)$ from five 2D projections using Cormack's method [6]. The experimental data (raw data) were deconvoluted with the Maximum Entropy algorithm [7] and the 3D densities were then calculated on planes perpendicular to the c-axis. The theoretical APW calculations embraced three different models for $\rho(p)$; $\rho^{IPM}(p)$, the density within the independent particle model (IPM), $\rho^{LDA1}(p)$, IPM plus a state-independent local-density enhancement factor and $\rho^{LDA2}(p)$, IPM plus a state-dependent, local-density enhancement factor (further details are given in [8]). $\rho^{e}(p)$, the electron momentum density, was also computed.

The theoretical densities were calculated along the directions $\Gamma M$ and $\Gamma K$ on six parallel planes $P_i$ perpendicular to the hexagonal axis, ½ $\Gamma A$ = 1.06 [mrad] apart (1 [mrad] = 0.137 [a.u]$^{-1}$). Since $E(\mathbf{k}) = E(\mathbf{k} + \mathbf{G})$, planes $\Gamma MK$ ($P_1$) and $\Gamma'M'K'$ ($P_5$) have the same energy bands (as do $P_2$ and $P_6$). However, in $\rho(p)$, symmetry selection rules [9] restrict the contributions from bands of a particular symmetry to particular parts of the momentum space $\mathbf{p} = \hbar(\mathbf{k} + \mathbf{G})$. Thus, in the extended zone scheme $\rho(p)$ contains contributions from all the occupied bands. But, the two full valence bands (see Fig. 1) have their main contribution to $\rho(p)$ within the two first Brillouin zones up to the plane $\Gamma MK$ (the first band between $\Gamma MK$ and $ALH$ and the second between $ALH$ and $\Gamma' M'K'$). The third band has a *neck* of holes around the $\Gamma$ point: this feature gets visible between the $ALH$ and $\Gamma' M'K'$ planes and becomes more and more dominant after crossing $\Gamma' M'K'$. The fourth band has its dominant contribution on the $\Gamma MK$ plane above the first Brillouin zone.

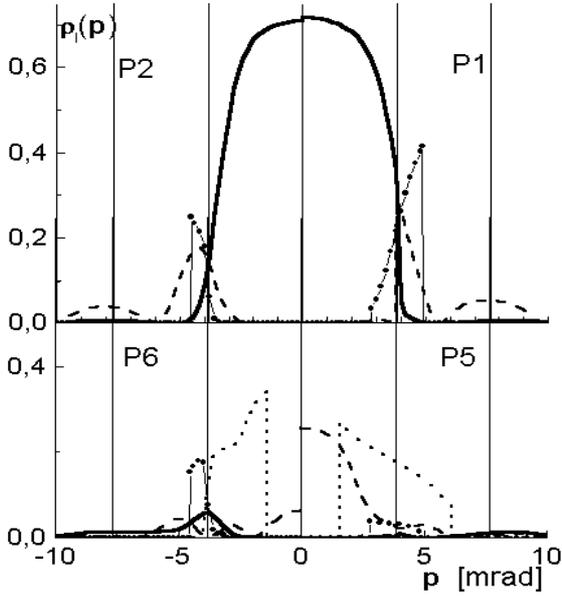

Fig. 1
Theoretical densities $\rho_l(p)$ in Y along $\Gamma M$ and parallel directions. Contributions from different valence bands ($l = 1, 2, 3, 4$) are marked by full, broken, dotted lines and dots, respectively. The thin perpendicular lines show the Brillouin zone boundaries at $\Gamma M$ and $2\Gamma M$ ($\Gamma M = 3.84$ mrad).

In order to compare theory with experiment we used the prescription $\rho(p) = \Sigma_l a_l \rho_l(p) + \rho_{core}(p)$ with constants $a_l$ which gave the best fit to the reconstructed experimental densities on plane $P_1$. The same $a_l$ were next applied to the theoretical results for all planes $P_i$. Fig. 2 shows the result. It is evident that the reconstructed density is essentially different from the electron density EMD. The best agreement between theory and experiment is found with $\rho^{LDA1}(p)$ but those LDA results are very similar to the IPM result and that similarity holds throughout the whole $p$ space [8]). Our new results [10] show that also the BML theory [11] fit reconstructed densities very well.

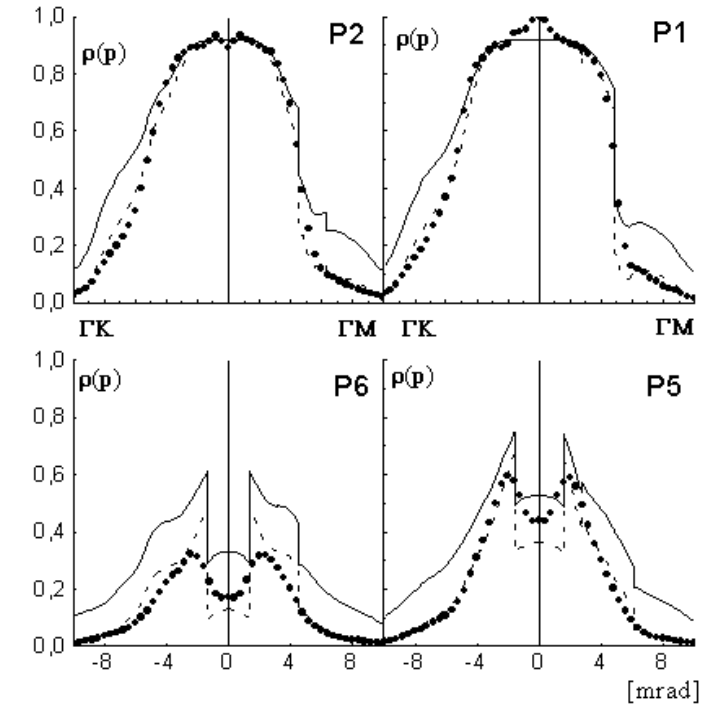

Fig. 2
Two theoretical models $\rho^e(p)$ (EMD) and state independent $\rho^{LDA1}(p)$ (full and dashed lines, respectively) compared with reconstructed densities (dots) along $\Gamma M$ and $\Gamma K$ on planes perpendicular to the hexagonal axis.

To obtain the shape of a FS one usually performs an LCW folding [5]. However, here, the redistribution of the different band contributions to different parts of the momentum space $p$, permits the ready identification of the individual FS sections. E.g. (see Figs. 1 and 2) on the plane $P_1$ and along $\Gamma M$ a rapid jump of $\rho(p)$ occurs for $p = 7.68$ ($2\Gamma M$) $- k_F = 4.88$ [mrad], where $k_F$ ($= 2.8$ [mrad]) is a Fermi radius of the 4$^{th}$ band around the $\Gamma$ point.

In Fig. 3 we show the FS of Y in an extended zone scheme deduced from Loucks results [1a]. On it we have marked those values of $p_F$ for which our theoretical densities $\rho(p)$ have jumps (compare Fig. 3 with Figs. 1 and 2) together with our estimates of the corresponding experimental values of

$p_F$, obtained by the method described in [7]. On the first two planes (4th band along $\Gamma M$) we obtain (in [a.u.]), respectively: 0.37 ± 0.01 (theory 0.38) and 0.43 ± 0.01 (theory 0.43) and in good agreement with Loucks. On the planes $P_5$ and $P_6$ (3rd band)) we obtained: $P_5$: $k_F(\Gamma M) = 0.16 \pm \Delta$ (theory 0.21), $k_F(\Gamma K) = 0.16 \pm \Delta$ (theory 0.22), $P_6$: $k_F(\Gamma M) = 0.23 \pm \Delta$ (theory 0.19) and $k_F(\Gamma K) = 0.23 \pm \Delta$ (theory 0.19), where $\Delta = 0.02$.

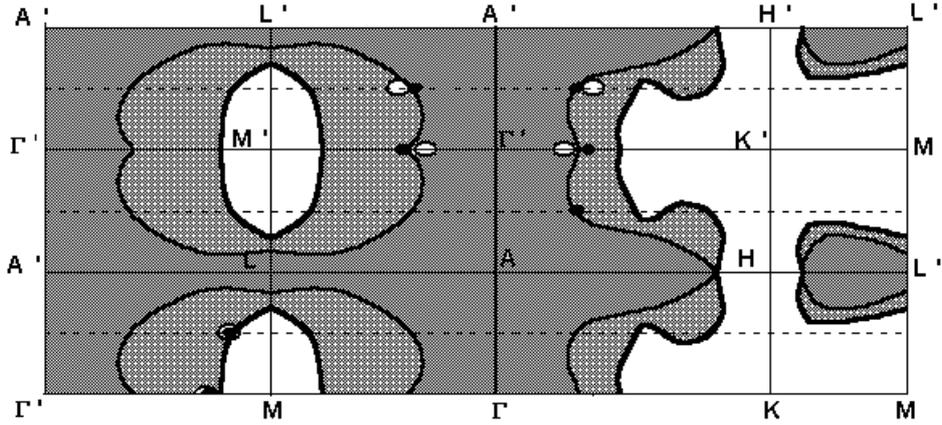

Fig. 3
Fermi surface of Y in the extended momentum space according to Loucks [1a]. Values of $p_F$, for which our theoretical densities have jumps are marked by solid circles. Corresponding values of $p_F$ obtained from reconstructed densities are marked by open circles.

The agreement between the experiment and both the theories (our and Loucks [1(a)] band structure results) is excellent for the 4th band along $\Gamma M$. In the 3rd band at $P_4$ and $P_6$ the experiment deviates more from the theories. However, here, $\rho(p)$ on *and* between the planes $P_4$ and $P_6$ changes more rapidly. This, together with the finite equipment resolution makes the estimate of the *cut-off* position susceptible to greater error.

**Acknowledgements.** We are grateful to Dr Ashraf Alam and Dr Stephen Dugdale for deconvoluting 2D ACAR yttrium experimental spectra.

For rapid communication (with G.Kontrym-Sznajd):
Tel.: (48-71) 34 350 21/131    Fax: (48-71) 441 029    e-mail: gsznajd@int.pan.wroc.pl